\documentclass[12pt]{iopart}
\usepackage{iopams}
\usepackage{epsf}
\usepackage{graphicx}

\begin{document}

\title{Inflation in a refined racetrack}

\author{Wen-Yu Wen}

\address{Department of Physics and Center for Theoretical Sciences,\\
National Taiwan University, Taipei 106, Taiwan}
\ead{steve.wen@gmail.com}

\begin{abstract}
In this note, we refine the racetrack inflation model constructed
in \cite{BlancoPillado:2004ns} by including the open string
modulus. This modulus encodes the embedding of our braneworld
inside some Calabi-Yau throat. We argue that in generic this open
string modulus dynamically runs with the inflaton field thanks to
its nonlinear coupling.  A full analysis becomes difficult because
the scalar potential changes progressively during the inflation
epoch. Nevertheless, by explicit construction we are still able to
build a realistic model through appropriate choices of the initial
conditions.
\end{abstract}



\maketitle


\section{Introduction and summary}
Several inflation models motivated by string theory have been
investigated so far\cite{inflation
model,Kachru,BlancoPillado:2004ns,BlancoPillado:2006he}. Provided
enormous landscape of vacua via various scenarios of string
compactification\cite{Denef}, they are equally interesting for the
reason that our universe might happen to be described by one of
them or a combination of some. String models of slow-roll
inflation fall into two major categories: In the scenario of brane
inlfation, the inflaton field is identified with the location of a
mobile D3-brane in a warped throat of the compactified manifold,
while in the scenario of moduli inflation the role of inflaton
field is played by one or more string moduli.  In particular, as a
simple example to the latter category, an inflation model was
developed using a racetrack potential inspired by KKLT
construction\cite{Kachru} within the context of type IIB string
theory\cite{BlancoPillado:2004ns}, and later refined by an
explicit construction\cite{Denef:2004dm} of specific orbifolded
Calabi-Yau manifold with two K\"{a}hler
moduli\cite{BlancoPillado:2006he}. One feature of this type of
inflation model is that the nonperturbative part of superpotential
has a form of double exponentials. With appropriate choices of the
parameters, this model can give rise to the desired slow-roll
inflation, which seems difficult to achieve in the original KKLT
model with single exponential.  Though both racetrack models gave
reasonable results consistent with the observational data, some
underlying assumptions may still be questionable.  One of them is
the assumption that all the moduli are fixed {\sl priorly} except
for some specific K\"{a}hler modulus, which acts as the inflaton
field.  This assumption may not be generically justified if some
moduli are coupled to each other too strongly to be integrated out
separately.  In this note, we would like to slightly relax the
above-mentioned assumption by allowing one open string modulus,
which encodes the embedding of our braneworld inside the
Calabi-Yau throat, freely run with the inflaton field. We will
only restrict our discussion based on the original racetrack
model\cite{BlancoPillado:2004ns} since the functional dependence
of open string modulus in the superpotential has been
known\cite{Baumann:2006th,DeWolfe:2007hd}. At the end, we find
that this modulus in generic takes nonzero expectation value and
therefore the racetrack potential would never be the same as that
in \cite{BlancoPillado:2004ns}. We have found our braneworld is
energetically pushed away from the tip of throat and the open
string modulus is settled at some nonzero value. Our result could
still be trusted even if deformation is applied to the conifold
singularity of tip, as long as deforming parameter is small enough
compared to the open string modulus. Moreover, contribution from
this modulus increases with the e-folding so that it cannot be
simply ignored during the period of inflation. A full analysis
becomes difficult because the scalar potential changes
progressively during the inflation epoch. Nevertheless by explicit
construction through appropriate choices of the initial conditions
in the moduli space, we are still able to build a realistic model
achieving enough e-foldings and consistent with the present day
observation.

This paper is organized in the following way.  First we briefly
review the original racetrack inflation model in the section 2. In
the section 3, we introduce the open string modulus and derive the
effective potential and discuss its vacua configuration.  Then in
the section 4, with appropriate choices of parameters and initial
conditions, we explicitly build a realistic racetrack model which
satisfies the slow-roll condition.  Some derivation of useful
formula are given in the Appendix.  In this note, we have set
Newton constant $G$ and Plank mass $M_{pl}$ to $1$ for
convenience.

We remark that different modification of racetrack inflation was
also discussed recently in \cite{deAlwis:2007qx,Brax:2007fe}.

\section{Brief review on racetrack inflation}
We begin with a brief review on derivation of racetrack potential
from the KKLT construction. We consider a compactification of
ten-dimensional type IIB string theory on Calabi-Yau manifold with
throats in the presence of three-form RR and NSNS fluxes
$H_{(3)}$\cite{Giddings:2001yu} and some stacks of D7-branes or
Euclidean D3-branes wrapping around four-cycles of the compact
space\cite{Kachru}. These background fluxes generate appropriate
potentials which fix the values of type IIB axion-dilaton field
and all the complex and K\"{a}hler moduli of Calabi-Yau space
except one, denoted as $T=X+iY$.  We may think of $X$ as the
volume of compactified space and $Y$ as the axion field.  Our
world may be seated on a stack of D3-branes inside one of the
throats. Now we consider a modified racetrack superpotential
inspired by KKLT,
\begin{equation}
W=W_0 + A_1e^{-a_1T} + A_2e^{-a_2T}.
\end{equation}
The constant term is obtained through
$W_0=\int{H_{(3)}\wedge\Omega}$ for the Calabi-Yau three form
$\Omega$. The exponential terms are obtained through gaugino
condensation nonperturbatively in a theory with a product gauge
group, in our case $SU(N_1)\times SU(N_2)$.  Coefficient
$a_i=2\pi/N_i$ and $N_i$ is the number of coincident D7 branes in
each stack\footnote{This is where the racetrack potential is
different from the original one used in the KKLT\cite{Kachru}.
There only one stack of D7-branes is considered.}. Coefficient
$A_i$'s come from the fact that gauge coupling is proportional to
the volume of compactified manifold and the latter is warped by
the presence of branes. They were fixed in the original racetrack
model but we will argue at this point in the next section.
Following KKLT, the scalar potential induced on our braneworld
composes of two parts
\begin{equation}
V=V_F+\delta V.
\end{equation}
We recall that $V_F$ is the F-term potential for the $N=1$
effective four-dimensional theory, namely,
\begin{equation}
V_F=e^K(K^{i\bar{j}}D_iW \overline{D_jW}-3|W|^2),
\end{equation}
provided that K{\"a}hler potential depends only on this K{\"a}hler
modulus or equivalently the volume of Calabi-Yau space, namely
\begin{equation}
K=-3\ln{(T+\bar{T})}=-3\ln{2X}.
\end{equation}
We remark that supersymmetric vacua should satisfy the following
conditions,
\begin{eqnarray}\label{susy_conditions}
D_iW=\partial_i W + \partial_i K W=0,
\end{eqnarray}
for each modulus, seen as a scalar field $\varphi_i$ in the
four-dimensional spacetime. Then the scalar potential is either
zero or negative, given by
\begin{equation}
V_F=-3e^{-K}|W|^2,
\end{equation}
corresponding to a flat or de Sitter spacetime.  However, we are
more interested in vacua with broken supersymetry.  The
supersymmetry breaking term $\delta V$ is induced by the presence
of anti-D3 branes, which can be placed at the end of throat and
stablized by the flux. Its contribution is positive definite and
takes form as follows,
\begin{equation}
\delta V=\frac{E}{X^{2}}.
\end{equation}
Since we ask modulus $T$ plays the role of inflaton field in the
background of broken supersymmetry, it does not respect the
condition (\ref{susy_conditions}).  With appropriate choices of
parameters, the authors in \cite{BlancoPillado:2004ns} found
several vacua where a saddle point exists with two local minima
nearby.  The $55$ e-foldings is achieved when the inflaton field
slowly rolls away from its saddle point and inflation is
terminated when it is trapped in one of the minima.

\begin{figure}
  \centering
  \includegraphics[width=0.49\textwidth]{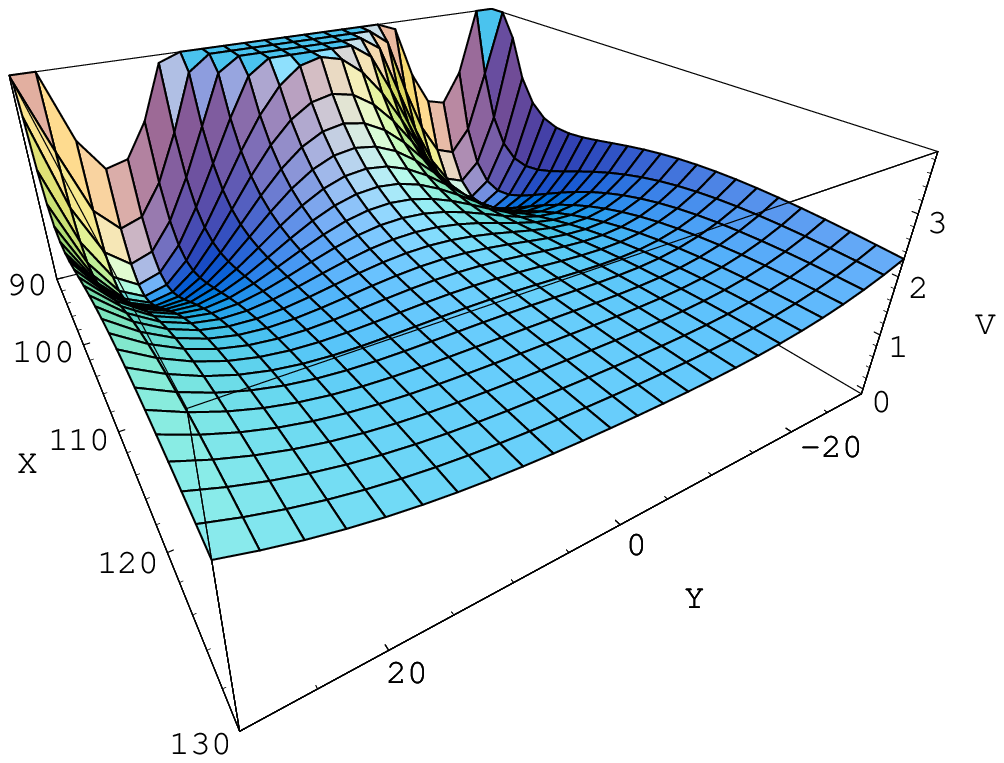}%
  \hfill%
  \includegraphics[width=0.49\textwidth]{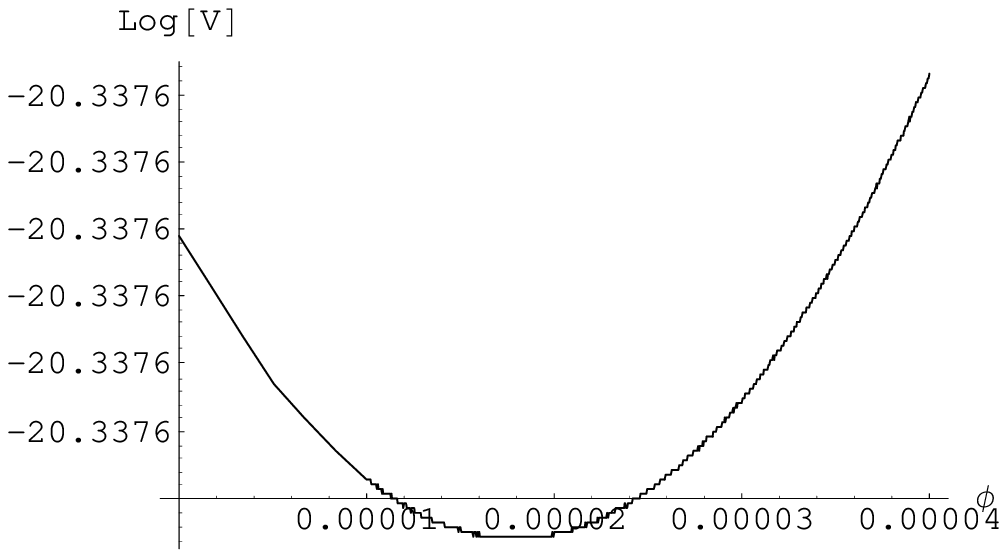}
  \caption{To the left, a snapshot of typical racetrack potential at local minimum of $\phi=1.79283\times 10^{-5}$.  The potential has a saddle point at $X=123.16,Y=0$ and two nearby local minima.
  To the right, it shows the plot of potential given $X=123.16,Y=0$ in the vicinity of the local minimum $\phi=1.79283\times 10^{-5}$.  The potential in both plots has been rescaled by $10^{16}$.}
\end{figure}

\section{Effective potential and vacua configuration in the refined racetrack}
In this section we would like to argue that the open string
modulus, denoted as $\phi$, which was priorly dropped off in
\cite{BlancoPillado:2004ns}, may not be consistently integrated
out at its supersymmetric vev thanks to the coupling with inflaton
$T$ through the exponential terms.  In this note, we may choose a
local complex patch $\{z^i\}$ near the end of throat and
$z\bar{z}=\phi$ is the embedding function of our braneworld inside
the throat. The argument is as follows: as long as the volume of
compactified manifold is changing during the inflation, the
relative location between our braneworld and stacks of D7-branes
are changing {\sl correspondingly}. Therefore, it is reasonable to
relax the assumption to allow modulus $\phi$ {\sl dynamically} run
with the K\"{a}hler modulus $T$.  In a simplest situation such as
the Ouyang embedding\cite{Ouyang:2003df}, D7-branes wrap around
the holomorphic four-cycles parameterized by one complex parameter
$\gamma_i^{-1}$.  Then this open string modulus appears in the
superpotential through $A_i$'s,
namely\cite{Baumann:2006th,DeWolfe:2007hd},
\begin{equation}
A_i(\phi)=\beta_i(1+\gamma_i\phi)^{1/N_i}.
\end{equation}
Assuming that K{\"a}hler potential has additional dependence on
the open string modulus $\phi$ as well as other messenger fields
$f$ and $\tilde{f}$ (fundamental rep of gauge group on each stack
of wrapped D7-branes.), such that a no-scale potential reads,
\begin{equation}
K=-3\ln{(T+\bar{T}-\phi\bar{\phi}-f\bar{f}-\tilde{f}\bar{\tilde{f}})}.
\end{equation}
Now we propose a more general form for superpotential, namely,
\begin{equation}
W=W_0 + A_1(\phi)e^{-a_1T} + A_2(\phi)e^{-a_2T}+m\phi f\tilde{f}.
\end{equation}

In particular, applying condition (\ref{susy_conditions}), we
obtain consistently truncated potential for $f=\tilde{f}=0$.
However, the modulus $\phi$ fails to be simply integrated out
provided the given superpotential and K\"{a}hler potential. To
simplify the discussion, we will make a further assumption that
the embedding only takes place in the real loci, i.e. both
$\gamma_i,\phi$ are real, though sometimes we still keep notation
of $\bar{\phi}$.  Then the scalar potential is again composed of
two parts,
\begin{equation}
V=V_F(X,Y,\phi)+\delta V(X).
\end{equation}
The effective potential $V_F$ contains two parts, whose explicit
forms are given in the Appendix. We have denoted $V_F^0$(see
equation (\ref{V0})) as a reduction of full potential $V_F$ for
$\phi=0$, which takes the same form as in
\cite{BlancoPillado:2004ns}. However, the other component, denoted
as $V_F^1$(see equation (\ref{V1})), was not included in
\cite{BlancoPillado:2004ns} and encodes most contribution of open
string modulus to the full potential. We here also choose a
specific setup suitable for racetrack inflation as found in
\cite{BlancoPillado:2004ns} out of many other possibilities, say
\begin{eqnarray}
&&\beta_1=\frac{1}{50},\qquad \beta_2=-\frac{35}{1000},\qquad
a_1=\frac{2\pi}{100},\qquad a_2=\frac{2\pi}{90},\nonumber\\
&&W_0=-\frac{1}{25000},\qquad E=4.14668\times 10^{-12}
\end{eqnarray}
with additional parameters of our choice,
\begin{equation}
\gamma_1=10^{-6},\gamma_2=-5.71429\times 10^{-7}.
\end{equation}
This choice respects the relation
$\gamma_1\beta_1=\gamma_2\beta_2$, which assures $D_{\phi}W=0$ at
expectation value $\phi=0$ in case $N_1=N_2$, though there is no
typical reason to do so in generic situation. For generic $N_1\neq
N_2$, $\phi=0$ is only achievable via $X\to \infty$, meaning that
this modulus is not stablized and the internal space becomes
decompactified. We also remark that all the scaling symmetries as
mentioned in \cite{BlancoPillado:2004ns}, i.e. $\gamma$ and
$\mu$-transformations, are broken for non-trivial dependence of
$A_i$ on $\phi$. However, the discrete symmetry along $Y$
direction is still preserved, i.e. $Y\sim Y+n L(N_1,N_2)$ for
integer $n$, where function $L(a,b)$ is the least common
multiplier of two integers $a$ and $b$. With these chosen
parameters, we have a saddle point at
\begin{equation}
X_s=123.216,\qquad Y_s=0,\qquad \phi_s=1.79283\times 10^{-5}.
\end{equation}
and nearby minima locate at
\begin{eqnarray}
&&X_m=93.7193,\qquad Y_m=\pm 23.1406,\qquad
\phi_m=5.78999.\nonumber\\
&&X_m=93.7193,\qquad Y_m=\pm 23.1407,\qquad \phi_m=-5.78999.
\end{eqnarray}
In the Figure 1, we plot the scalar potential at $\phi=\phi_s$. We
remark several properties of this moduli space. First notice that
$\phi\ne 0$ is expected in those specific locations, therefore the
racetrack potential in \cite{BlancoPillado:2004ns} is {\sl not}
energetically favored in our refined model. During the inflation,
we have found our braneworld, initially at $\phi=\phi_s$, is
energetically pushed away from the tip of throat and settled at
$\phi=\phi_m$ for the open string modulus.  Our result could still
be trusted even if deformation is applied to the conifold
singularity of tip, as long as deforming parameter is small enough
compared to $\phi_s$ and $\gamma_i^{-1}$.

Secondly, $V_{,yy}<0$ at the saddle point and one may guess that
the initial motion of slow-roll is along Y-direction at least in
case of no initial speed. We will see that this may not be true in
our refined model.

At last, the contribution from $V_F^1$ becomes significant at
large value of $\phi$; to see that we found $|V_F^1/V_F^0|$ varies
from $10^{-11}$ to $0.28$ for inflaton moves from the saddle point
to one of the minima, while $\delta V$ is about the same order as
$V_F^0$. The saddle point implies a de Sitter space with $V\simeq
1.65489\times 10^{-16}$ and the minima are anti-de Sitter space
with $V\simeq -1.148\times 10^{-16}$. Both have magnitude of order
$10^{-16}$. If we increase $E$ up to $5.15815\times 10^{-12}$ or
more in order to have Minkovski or de Sitter vacua at the
minima\footnote{With this new $E$, minima shift to
$(X_m,Y_m,\phi_m)=(93.9764,\pm 23.1254,\pm 5.7632)$.}, then the
saddle point disappears and the slow-roll may not take place.

\begin{figure}
  \centering
  \includegraphics[width=0.49\textwidth]{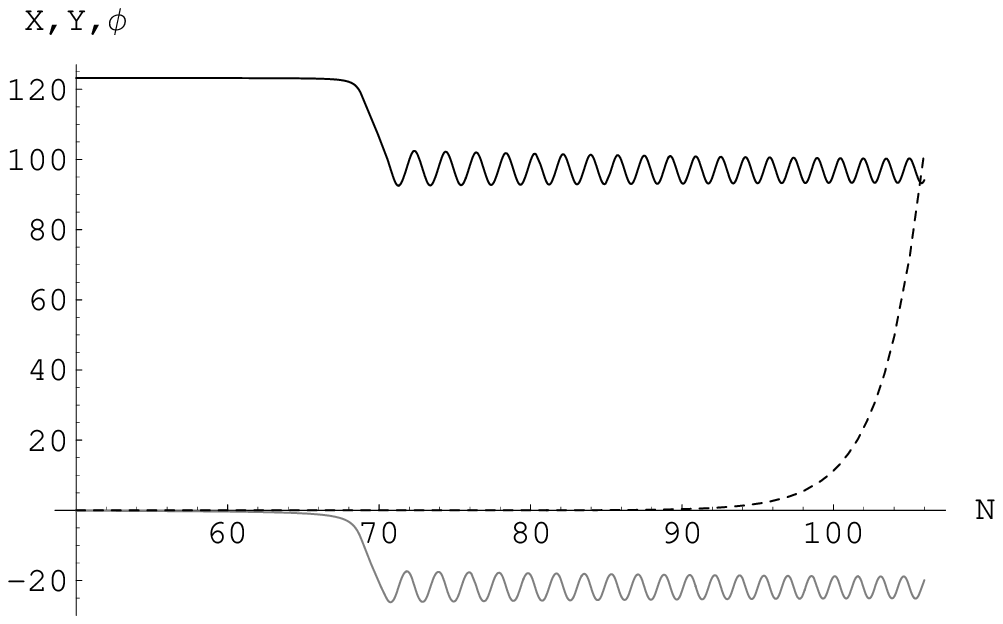}%
  \hfill%
  \includegraphics[width=0.49\textwidth]{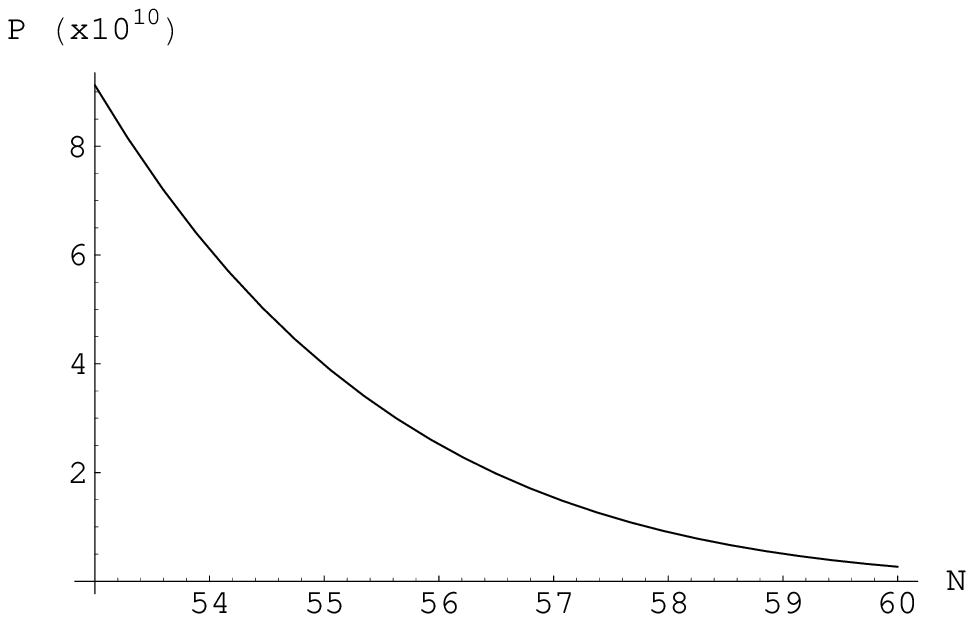}
  \caption{To the left, the plot shows each moduli evolves with
  e-folding $N$.  The black solid line stands for modulus $X$, gray
  solid line for $Y$ and dashed line for $\phi$.
  To the right, it shows the power spectrum of scalar density perturbation evaluated at different values of $N$.
  We remark that $P\simeq 4\times 10^{-10}$ at $N=55$.}
\end{figure}

\section{Slow-roll inlfation}
In this section, we would like to discuss the condition for
slow-roll inflation.  We recall that the kinetic term of scalar
fields are given in term of the K\"{a}hler potential (see equation
(\ref{metric})), that is,
\begin{eqnarray}
{\cal L}_{kin}&&=\frac{1}{2}{\cal
G}_{ij}\partial_{\mu}\varphi^i\partial^{\mu}\varphi^j\nonumber\\
&&=\frac{3}{\Delta^2}(\partial_{\mu}
X\partial^{\mu}X+\partial_{\mu}Y\partial^{\mu}Y)+\frac{6(\Delta-X)}{\Delta^2}\partial_\mu
\phi\partial^\mu\phi.
\end{eqnarray}

For scalar fields having non-canonical kinetic terms, we have
slow-roll parameters in general forms,
\begin{eqnarray}
&&\epsilon=\frac{1}{2}(\frac{g^{ij}V_{,i}V_{,j}}{V^2}),\nonumber\\
&&N^{i}{}_j=[\frac{g^{ik}(V_{,kj}-\Gamma^l_{kj}V_{,l})}{V}],
\end{eqnarray}
where $\eta$ is defined as the most negative eigenvalue of the
matrix $N^i{}_j$.  At the saddle point, $\epsilon$ is trivially
zero and $\eta=-0.0384854$ for parameters of our choice.

The evolution of scalar fields is govern by the following
equations of motion,
\begin{eqnarray}\label{eom}
&&\ddot{\varphi^i}+3H\dot{\varphi^i}+\Gamma^i_{jk}\dot{\varphi^j}\dot{\varphi^k}+g^{ij}\frac{\partial
V}{\partial\varphi^j}=0,\nonumber\\
&&3H^2=\frac{1}{2}{\cal G}_{ij}\dot{\varphi^i}\dot{\varphi^j}+V,
\end{eqnarray}
where the Hubble {\sl constant} $H\equiv \dot{a}/a$.  We have
listed equations of motion for each scalar $\varphi^i=X,Y,\phi$ in
the Appendix\footnote{Throughout this paper, we use the notation
$\dot{\varphi}=\partial_t \varphi$ and $\varphi'=\partial_N
\varphi$, where e-folding $N\equiv\ln{a(t)}$.}. The term linear to
$\dot{\varphi}$ plays the role of friction, while the gradient of
potential gives rise to the conservative force. We remark that the
nontrivial Levi-Civita connections defined on the moduli space
encode the complicate couplings among those scalars fields and we
have them listed in the Appendix too.

Assuming that inflation starts at rest at the saddle point, we
find that before inflation ends at $\epsilon=1$, we only have
about $47$ e-foldings, which is too small for our observational
universe. However, with fine-tuned initial speed $X_s'=-0.02$, we
are able to achieve in total about $106$ e-foldings, as shown in
the Figure 2. We also have power spectrum of scalar density
perturbation $P(k)\simeq 4\times 10^{-10}$ at $55$ e-foldings,
qualifying the COBE normalization. Provided
$\epsilon=2.80795\times 10^{-10}$ and $\eta=-0.0235525$ at the
$55$ e-foldings, one obtains the spectral index,
\begin{equation}
n(k)\simeq 1-6\epsilon + 2\eta = 0.952895,
\end{equation}
which agrees well with the observational data.  We remark that the
initial motion of inflaton field with zero initial speed is in
fact along $\phi$-direction, instead of $Y$-direction as observed
in \cite{BlancoPillado:2004ns}.  The inflaton field with $X'_s\neq
0$ makes additional detour along $X$-direction.  We plot both
cases in the Figure 3 for the first few e-foldings.

We argue that there seems no criteria to assign any particular
initial condition for inflaton fields, provided no sound knowledge
of probability measure in the moduli phase space. Therefore there
seems no objection to assign some initial speed for inflaton
fields as we just did.  It might stand a good chance to find
another set of parameters with inflaton fields of zero initial
speed and achieving enough e-foldings, but that could be an
exhausting search without some clever guiding principles.  We will
leave this issue for further study.

\begin{figure}
  \centering
  \includegraphics[width=0.49\textwidth]{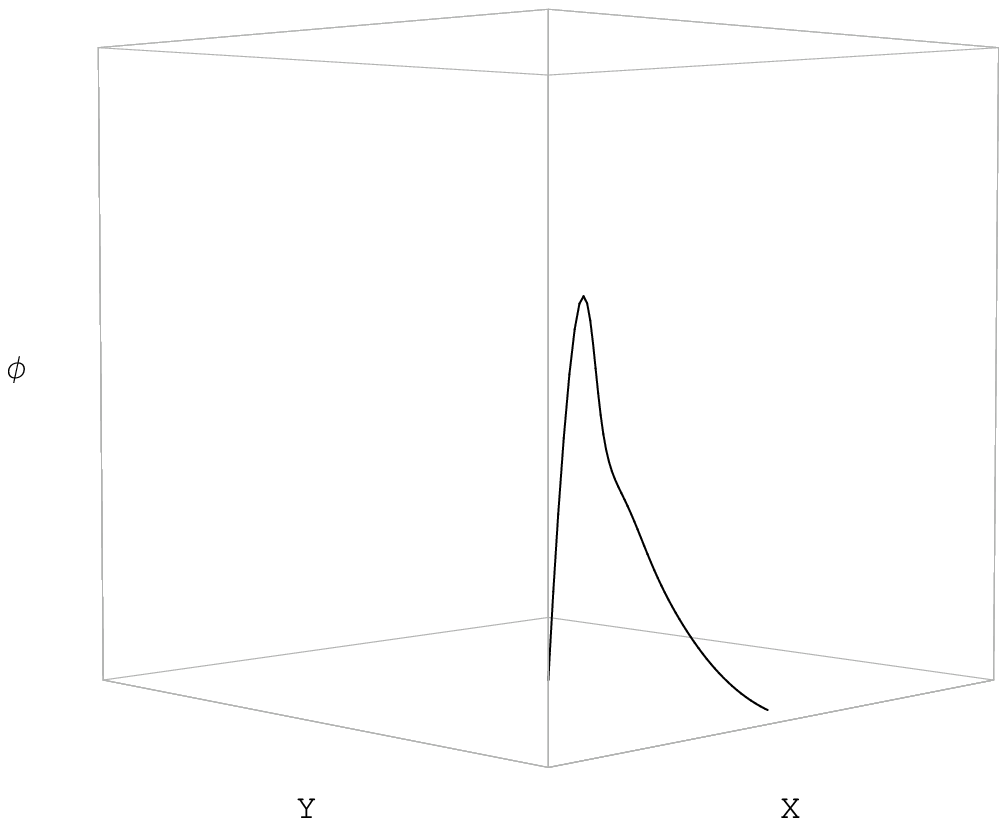}%
  \hfill%
  \includegraphics[width=0.49\textwidth]{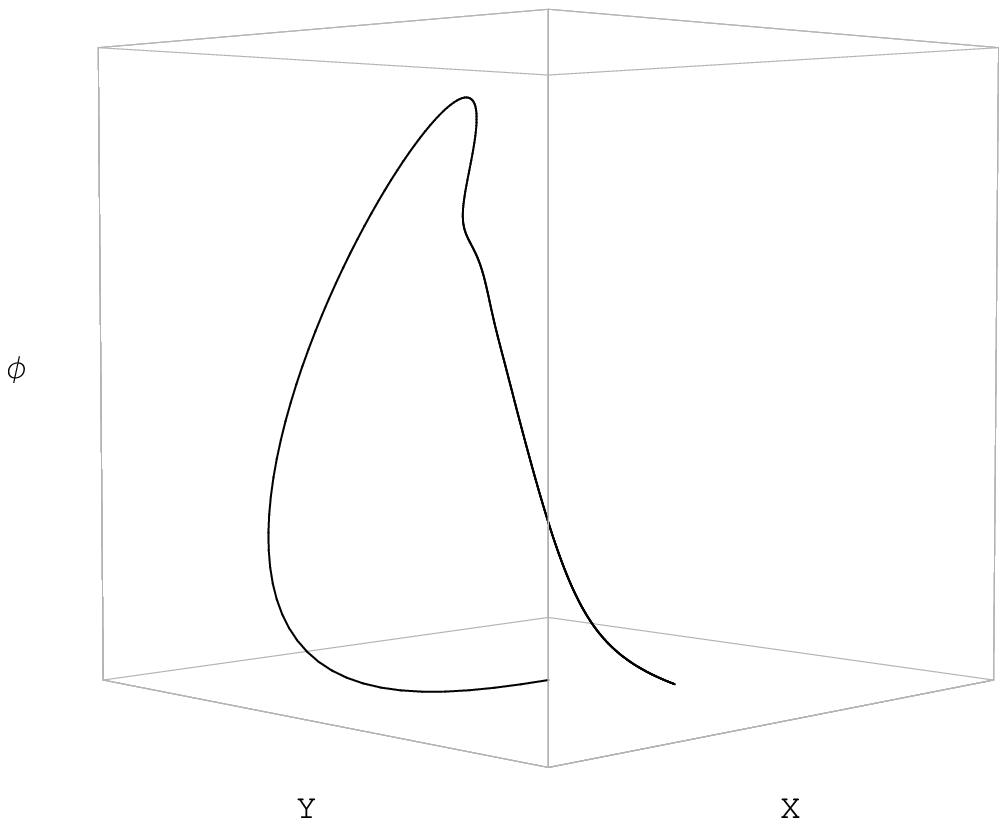}
  \caption{To the left, inflaton starts at rest, first moves towards
  $\phi$-direction and then $Y$-direction.
  To the right, infaton with initial speed $X'_s=-0.02$ first makes
  additional detour along $X$-direction.  The typical scale of each axis $X,Y,\phi$ is about $0.01,0.01$, and $10^{-9}$.}
\end{figure}

\ack
I am grateful to the invitation of CosPA '07 conference to
present this work at its early stage.  I would like to thank P.-M.
Ho for useful discussion.  The author is supported in part by the
Taiwan's National Science Council under Grant No.
NSC96-2811-M-002-018 and NSC96-2119-M-002-001.

\section*{Appendix}
 Here we have listed the explicit forms of scalar
 potential and equations of motion.  First a collection of useful
 formula for K\"{a}hler metrics and derivatives,
\begin{eqnarray}\label{metric}
&&K_{T\bar{T}}=\frac{1}{2}{\cal G}_{XX}=\frac{1}{2}{\cal
G}_{YY}=\frac{3}{\Delta^2},\qquad
K_{\phi\bar{\phi}}=\frac{1}{2}{\cal G}_{\phi\phi}=\frac{6(\Delta-X)}{\Delta}\nonumber\\
&&D_TW=-a_1A_1e^{-a_1T}-a_2A_2e^{-a_2T}-\frac{3}{\Delta}W,\nonumber\\
&&D_\phi W =
A_i'e^{-a_1T}+A_2'e^{-a_2T}+\frac{3\bar{\phi}}{\Delta}W,\nonumber\\
&&\Delta=2X-|\phi|^2,\qquad.
\end{eqnarray}
The scalar potential then reads,
\begin{eqnarray}
V&=&V^0_F+V^1_F+\delta V,\nonumber\\
\label{V0}V^0_F&=&\frac{1}{3\Delta}\biggl[
a_1(a_1+\frac{6}{\Delta})A_1^2 e^{-2a_1X}+a_2
(a_2+\frac{6}{\Delta})A_2^2e^{-2a_2X}\nonumber\\
&&+(a_1a_2+\frac{3}{\Delta}a_1+\frac{3}{\Delta}a_2)A_1A_2e^{-(a_1+a_2)X}2\cos{(a_1-a_2)Y}\nonumber\\
&&+\frac{3}{\Delta}a_1W_0A_1e^{-a_1X}2\cos{a_1Y}+\frac{3}{\Delta}a_2W_0A_2e^{-a_2X}2\cos{a_2Y}\biggr]\\
\label{V1}V^1_F&=&\frac{1}{6\Delta(\Delta-X)}\biggl[
A_1'^2e^{-2a_1X}+A_2'^2e^{-2a_2X}+\frac{9|\phi|^2}{\Delta^2}|W|^2\nonumber\\
&&+A_1'A_2'e^{-(a_1+a_2)X}2\cos{(a_1-a_2)Y}+\frac{3}{\Delta}(A_1'\square_1+A_2'\square_2)
\biggr],\\
\delta V&=&\frac{E}{X^2},\\
&&\square_i=-2\phi e^{-a_iX}(W_1\cos{a_iY}+W_2\sin{a_iY}),
\end{eqnarray}
where $T=X+iY$ and $W=W_1+iW_2$.

The nontrivial Levi-Civita connections in the moduli space are
given as follows:
\begin{eqnarray}
&&\Gamma^X_{XX}=-\Gamma^X_{YY}=-\frac{2}{\Delta}, \qquad
\Gamma^X_{\phi\phi}=\frac{\Delta-2\phi^2}{\Delta}, \qquad
\Gamma^X_{X\phi}=\frac{2\phi}{\Delta},\nonumber\\
&&\Gamma^Y_{XY}=-\frac{2}{\Delta},\qquad
\Gamma^Y_{Y\phi}=\frac{2\phi}{\Delta},\\
&&\Gamma^{\phi}_{\phi\phi}=-\frac{|\phi|^2\phi}{\Delta(\Delta-X)},\qquad
\Gamma^{\phi}_{XX}=-\frac{\phi}{\Delta(\Delta-X)},\nonumber\\
&&\Gamma^{\phi}_{X\phi}=\frac{2|\phi|^2-\Delta}{2\Delta(\Delta-X)},\qquad
\Gamma^{\phi}_{YY}=\frac{\phi}{\Delta(\Delta-X)}.\nonumber
\end{eqnarray}

Applying the equation (\ref{eom}) for each scalar fields, we have
equations of motion
\begin{eqnarray}
&&X''=-\Omega(3X'+\frac{\Delta^2}{2}\frac{V_{,X}}{V})+\frac{1}{\Delta}(2X'^2-2Y'^2-(\Delta-2\phi^2)\phi'^2-4\phi
X'\phi')\nonumber\\
&&Y''=-\Omega(3Y'+\frac{\Delta^2}{2}\frac{V_{,Y}}{V})+\frac{4}{\Delta}(X'Y'-\phi\phi'Y')\nonumber\\
&&\phi''=-\Omega(3\phi'+\frac{\Delta^2}{4(\Delta-X)}\frac{V_{,\phi}}{V})+\frac{1}{\Delta(\Delta-X)}[\phi(X'^2+Y'^2)+\phi^3\phi'^2+(\Delta-2\phi^2)X'\phi'],\nonumber\\
&&\Omega = 1-\frac{X'^2+Y'^2+2(\Delta-X)\phi'^2}{\Delta^2}.
\end{eqnarray}


\section*{References}

\begin{thebibliography}{99}


\bibitem{inflation model}
  G.~R.~Dvali and S.~H.~H.~Tye,
  Phys.\ Lett.\  B {\bf 450}, 72 (1999)
  [arXiv:hep-ph/9812483].

  C.~P.~Burgess, M.~Majumdar, D.~Nolte, F.~Quevedo, G.~Rajesh and R.~J.~Zhang,
  JHEP {\bf 0107}, 047 (2001)
  [arXiv:hep-th/0105204].

  G.~R.~Dvali, Q.~Shafi and S.~Solganik,
  arXiv:hep-th/0105203.

  J.~Garcia-Bellido, R.~Rabadan and F.~Zamora,
  JHEP {\bf 0201}, 036 (2002)
  [arXiv:hep-th/0112147].

  N.~T.~Jones, H.~Stoica and S.~H.~H.~Tye,
  JHEP {\bf 0207}, 051 (2002)
  [arXiv:hep-th/0203163].

  M.~Gomez-Reino and I.~Zavala,
  JHEP {\bf 0209}, 020 (2002)
  [arXiv:hep-th/0207278].


  C.~P.~Burgess, J.~M.~Cline, H.~Stoica and F.~Quevedo,
  JHEP {\bf 0409}, 033 (2004)
  [arXiv:hep-th/0403119].

  J.~M.~Cline and H.~Stoica,
  Phys.\ Rev.\  D {\bf 72}, 126004 (2005)
  [arXiv:hep-th/0508029].

  J.~P.~Hsu, R.~Kallosh and S.~Prokushkin,
  JCAP {\bf 0312}, 009 (2003)
  [arXiv:hep-th/0311077].

  F.~Koyama, Y.~Tachikawa and T.~Watari,
  Phys.\ Rev.\  D {\bf 69}, 106001 (2004)
  [Erratum-ibid.\  D {\bf 70}, 129907 (2004)]
  [arXiv:hep-th/0311191].

  H.~Firouzjahi and S.~H.~H.~Tye,
  Phys.\ Lett.\  B {\bf 584}, 147 (2004)
  [arXiv:hep-th/0312020].

  J.~P.~Hsu and R.~Kallosh,
  JHEP {\bf 0404}, 042 (2004)
  [arXiv:hep-th/0402047].

  E.~Silverstein and D.~Tong,
  Phys.\ Rev.\  D {\bf 70}, 103505 (2004)
  [arXiv:hep-th/0310221].

  R.~Kallosh and S.~Prokushkin,
  arXiv:hep-th/0403060.


  M.~Alishahiha, E.~Silverstein and D.~Tong,
  Phys.\ Rev.\  D {\bf 70}, 123505 (2004)
  [arXiv:hep-th/0404084].

  X.~Chen,
  JHEP {\bf 0508}, 045 (2005)
  [arXiv:hep-th/0501184].

  X.~Chen,
  Phys.\ Rev.\  D {\bf 72}, 123518 (2005)
  [arXiv:astro-ph/0507053].

  D.~Cremades, F.~Quevedo and A.~Sinha,
  JHEP {\bf 0510}, 106 (2005)
  [arXiv:hep-th/0505252].

  C.~P.~Burgess, P.~Martineau, F.~Quevedo, G.~Rajesh and R.~J.~Zhang,
  JHEP {\bf 0203}, 052 (2002)
  [arXiv:hep-th/0111025].

  Z.~Lalak, G.~G.~Ross and S.~Sarkar,
  Nucl.\ Phys.\  B {\bf 766}, 1 (2007)
  [arXiv:hep-th/0503178].

  J.~P.~Conlon and F.~Quevedo,
  JHEP {\bf 0601}, 146 (2006)
  [arXiv:hep-th/0509012].

  O.~DeWolfe, S.~Kachru and H.~L.~Verlinde,
  JHEP {\bf 0405}, 017 (2004)
  [arXiv:hep-th/0403123].

  N.~Iizuka and S.~P.~Trivedi,
  Phys.\ Rev.\  D {\bf 70}, 043519 (2004)
  [arXiv:hep-th/0403203].

\bibitem{Kachru}
  S.~Kachru, R.~Kallosh, A.~Linde and S.~P.~Trivedi,
  Phys.\ Rev.\  D {\bf 68}, 046005 (2003)
  [arXiv:hep-th/0301240].

  S.~Kachru, R.~Kallosh, A.~Linde, J.~M.~Maldacena, L.~P.~McAllister and S.~P.~Trivedi,
  JCAP {\bf 0310}, 013 (2003)
  [arXiv:hep-th/0308055].


\bibitem{BlancoPillado:2004ns}
  J.~J.~Blanco-Pillado {\it et al.},
  JHEP {\bf 0411}, 063 (2004)
  [arXiv:hep-th/0406230].

\bibitem{BlancoPillado:2006he}
  J.~J.~Blanco-Pillado {\it et al.},
  JHEP {\bf 0609}, 002 (2006)
  [arXiv:hep-th/0603129].

\bibitem{Denef}
  F.~Denef and M.~R.~Douglas,
  JHEP {\bf 0405}, 072 (2004)
  [arXiv:hep-th/0404116].

  F.~Denef and M.~R.~Douglas,
  JHEP {\bf 0503}, 061 (2005)
  [arXiv:hep-th/0411183].

  B.~S.~Acharya, F.~Denef and R.~Valandro,
  JHEP {\bf 0506}, 056 (2005)
  [arXiv:hep-th/0502060].

\bibitem{Denef:2004dm}
  F.~Denef, M.~R.~Douglas and B.~Florea,
  JHEP {\bf 0406}, 034 (2004)
  [arXiv:hep-th/0404257].

\bibitem{Baumann:2006th}
  D.~Baumann, A.~Dymarsky, I.~R.~Klebanov, J.~M.~Maldacena, L.~P.~McAllister and A.~Murugan,
  JHEP {\bf 0611}, 031 (2006)
  [arXiv:hep-th/0607050].

\bibitem{DeWolfe:2007hd}
  O.~DeWolfe, L.~McAllister, G.~Shiu and B.~Underwood,
  JHEP {\bf 0709}, 121 (2007)
  [arXiv:hep-th/0703088].

\bibitem{deAlwis:2007qx}
  S.~P.~de Alwis,
  Phys.\ Rev.\  D {\bf 76}, 086001 (2007)
  [arXiv:hep-th/0703247].

\bibitem{Brax:2007fe}
  P.~Brax, A.~C.~Davis, S.~C.~Davis, R.~Jeannerot and M.~Postma,
  arXiv:0710.4876 [hep-th].

\bibitem{Giddings:2001yu}
  S.~B.~Giddings, S.~Kachru and J.~Polchinski,
  Phys.\ Rev.\  D {\bf 66}, 106006 (2002)
  [arXiv:hep-th/0105097].

\bibitem{Ouyang:2003df}
  P.~Ouyang,
  Nucl.\ Phys.\  B {\bf 699}, 207 (2004)
  [arXiv:hep-th/0311084].






\end{thebibliography}
\end{document}